\documentclass[conference]{IEEEtran}
\IEEEoverridecommandlockouts
\usepackage{cite}
\usepackage{amsmath,amssymb,amsfonts}
\usepackage{algorithm}
\usepackage{algorithmic}
\usepackage{graphicx}
\usepackage{textcomp}
\usepackage{xcolor}
\usepackage{comment}
\usepackage{booktabs}
\def\BibTeX{{\rm B\kern-.05em{\sc i\kern-.025em b}\kern-.08em
    T\kern-.1667em\lower.7ex\hbox{E}\kern-.125emX}}
\begin{document}

\title{SMC-AI: Scaling Monte Carlo Simulation to Four Trillion Atoms with AI Accelerators}


%
%
%
%
%
%

\author{\IEEEauthorblockN{Xianglin Liu$^{*\dagger}$ \thanks{*Co-first author} \thanks{$\dagger$Corresponding author}}
\IEEEauthorblockA{\textit{Pengcheng Laboratory} \\
Shenzhen, China \\
xianglinliu01@gmail.com}
\and
\IEEEauthorblockN{Kai Yang$^{*}$}
\IEEEauthorblockA{\textit{Pengcheng Laboratory} \\
Shenzhen, China \\
yangk@pcl.ac.cn}
\and
\IEEEauthorblockN{Fanli Zhou}
\IEEEauthorblockA{\textit{School of Computer and Artificial Intelligence} \\ 
\textit{Xiangnan University} \\
Chenzhou, China \\
flizhou@gmail.com}
\and
\IEEEauthorblockN{Yongxiang Liu}
\IEEEauthorblockA{\textit{Pengcheng Laboratory} \\
Shenzhen, China \\
liuyx@pcl.ac.cn}
\and
\IEEEauthorblockN{Hao Chen}
\IEEEauthorblockA{\textit{Pengcheng Laboratory} \\
Shenzhen, China \\
\textit{Southern University of Science and Technology} \\
Shenzhen, China \\
chenh10@pcl.ac.cn}
\and
\IEEEauthorblockN{Yijia Zhang}
\IEEEauthorblockA{\textit{Pengcheng Laboratory} \\
Shenzhen, China \\
zhangyj01@pcl.ac.cn}
\and
\IEEEauthorblockN{Dengdong Fan}
\IEEEauthorblockA{\textit{Pengcheng Laboratory} \\
Shenzhen, China \\
fandd@pcl.ac.cn}
\and
\IEEEauthorblockN{Wenbo Li}
\IEEEauthorblockA{\textit{Pengcheng Laboratory} \\
Shenzhen, China \\
liwb01@pcl.ac.cn}
\and
\IEEEauthorblockN{Bingqiang Wang}
\IEEEauthorblockA{\textit{Pengcheng Laboratory} \\
Shenzhen, China \\
wangbq@pcl.ac.cn}
\and
\IEEEauthorblockN{Shixun Zhang}
\IEEEauthorblockA{\textit{Pengcheng Laboratory} \\
Shenzhen, China \\
zhangshx@pcl.ac.cn}
\and
\IEEEauthorblockN{Pengxiang Xu$^\dagger$}
\IEEEauthorblockA{\textit{Pengcheng Laboratory} \\
Shenzhen, China \\
xupx@pcl.ac.cn}
\and
\IEEEauthorblockN{Yonghong Tian}
\IEEEauthorblockA{\textit{School of AI for Science} \\
\textit{Shenzhen Graduate School, Peking University} \\
Shenzhen, China \\
\textit{Pengcheng Laboratory} \\
Shenzhen, China \\
yhtian@pku.edu.cn}}


\maketitle

\begin{abstract}
The rapid advancement of deep learning is reshaping the hardware design landscape toward AI tasks, posing fundamental challenges for HPC workloads such as atomistic simulation. Here we present SMC-AI, a general algorithmic framework that extends the SMC-X method for efficient canonical Monte Carlo simulation on AI accelerators, including GPUs and NPUs, while maintaining extreme scalability. The implementation of SMC-AI on an NPU cluster reaches unprecedented performance, achieving MC simulation of 4 trillion atoms on 4096 NPU dies. This represents the largest ML-accelerated atomistic simulation reported, delivering 32× system size and 1.3× throughput than previous records, with a relatively small computational budget. Excellent strong and weak scaling efficiency are reached for both the NPU and GPU implementation. By decoupling ML models from simulation, SMC-AI creates an abstraction that facilitates integration and porting of diverse ML models, laying a foundation for the future development of scalable scientific software.
\end{abstract}


\section{Introduction}
The rapid advancement of deep learning has driven hardware design to prioritize AI workloads. A wide range of AI-specific accelerators has emerged, including Google TPU \cite{9351692}, Huawei Ascend NPU \cite{zuo2025servinglargelanguagemodels}, Cerebras Wafer-Scale Engine \cite{10793165}, and AWS Trainium \cite{10.1145/3698038.3698535}. At the same time, general-purpose processors, such as CPUs \cite{IntelSaphaireRapids} and GPUs \cite{10070122} increasingly incorporate AI-oriented features, including a larger number of tensor cores, low-precision arithmetic units, and bandwidth-optimized memory hierarchies. These designs favor high FLOPS and energy efficiency and provide opportunities for the reuse of highly optimized AI compute and interconnect infrastructure for high-performance computing (HPC) workloads.
Together with the rapidly increasing cost of building flagship computing systems, it seems increasingly attractive, and sometimes inevitable, for scientific applications to run on hardware primarily designed for AI workloads. However, the  architectural choices of AI accelerators often diverge from the needs of HPC applications, which generally have significantly different arithmetic intensity and memory access patterns\cite{JackNaturePhysics}. Exploring strategies to overcome these challenges is, therefore, critical to enabling the next generation of HPC applications in the current age of AI.

Among various HPC applications, a particularly important subfield is the use of ML surrogates to replace, and dramatically accelerate, the evaluation of quantum mechanics in atomistic simulation. A most well-known example is
machine-learning interatomic potentials (MLIPs) \cite{2025Nature} to accelerate first principles molecular dynamics. This method of ML-accelerated atomistic simulation has revolutionized molecular and material modeling \cite{MLIP-Central-NCS} by combining high computational efficiency with near first-principles accuracy, offering performance comparable to classical empirical models while approaching the precision of first-principles methods such as density functional theory (DFT) \cite{Zhang_2025}. This capability enables atomistic simulation to serve as a “computational microscope” \cite{ComputationalMicroscopeStructure, annurev:/content/journals/10.1146/annurev-biophys-042910-155245, zhou2026computationalmicroscopechemicalorderdisorder}, allowing scientists to probe mesoscale phenomena at atomic resolution. Representative examples include virus capsid assembly \cite{10.1145/3581784.3627041}, plastic deformation in alloys \cite{Yin2021}, and thermal transport in materials \cite{ZHOU2025116568}.

Currently, ML surrogates are typically integrated into atomistic simulations via heterogeneous computation: ML model inference runs on GPUs to harness high FLOPS, while the simulation runs on CPUs for atomistic position storage, updating, and input feature calculation. Some softwares adopt relatively coarse-grained coupling between CPU and GPU executions, which is a strategy to favor modular design, and makes the incorporation of MLIPs straightforward, as exemplified by the LAMMPS package's porting of SNAP \cite{10.1145/3458817.3487400}, DeePMD \cite{10.5555/3433701.3433707}, and Allegro \cite{10.1145/3581784.3627041} models.
However, such a benefit typically comes at the price of  computational efficiency. For molecular dynamics (MD), there are recently developed software, such as GPUMD \cite{FAN201710}, that adopts a more aggressive optimization strategy of offloading most work to GPUs to achieve higher speed, as shown in Tab.~\ref{Tab:Performance}. Similarly, for Monte Carlo (MC), the recently proposed SMC-X method \cite{liu2025revealingnanostructureshighentropyalloys, doi:10.1021/acs.jctc.5c01614} attains orders-of-magnitude speedup over CPU implementations \cite{sadigh2012scalable}, due to both algorithmic innovations and fully executing ML inference and MC simulation on GPUs. Nevertheless, the original SMC-X algorithm is designed for general-purpose chips such as CPUs and GPUs, and thus faces fundamental challenges when ported to AI-specific chips. Moreover, the tight coupling design in SMC-X, while rendering it extremely efficient, makes it cumbersome to integrate with rapidly evolving ML models, which are becoming increasingly complex due to various physics-embedding schemes \cite{NequIP_NC}.

The main contributions are the following. First, we introduce the SMC-AI algorithm to address the above challenges. As a variant of the SMC-X algorithm, SMC-AI is designed to take advantage of the computational capability of AI-centric hardware architecture while retaining the excellent performance of SMC-X as much as possible. Second, through algorithm redesign, as well as a series of tailored optimization techniques, we demonstrate that SMC-AI scales efficiently on a large-scale NPU cluster, achieving a record-breaking simulation of 4 trillion atoms across 4,096 NPU dies. This yields a 32× increase in spatial scale and a 1.3× improvement in throughput over previous state-of-the-art ML-accelerated atomistic simulations, all with a much smaller computational budget, as shown in Tab.~\ref{Tab:Performance}. 
We further evaluate both strong and weak scaling of SMC-AI on NPUs and GPUs, and provide a performance analysis of the hotspot kernel, focusing on arithmetic intensity. Finally, using the proposed MLPNet as a representative example, we show that SMC-AI introduces a clean abstraction between ML models and the Monte Carlo core, providing a flexible foundation for integrating diverse models in future developments.

\begin{table*}
  \caption{Comparison of the performance of atomistic simulations (MC and MD) achieving DFT-level accuracy (10 meV/atom). 
  }
  \label{Tab:Performance}
\begin{tabular}{c c c c c c c c }
\toprule
\textbf{Application}  & \textbf{Potential} & \textbf{System} & \textbf{Chips} & \textbf{Machine} & \textbf{Atoms} & \textbf{Throughput (TP)}  \\
 &  &  &  &  & \textbf{(billion)} & \textbf{ (billion atom$\cdot$step/s)}  \\
\midrule
WL-LSMS \cite{10.1145/1654059.1654125}  & DFT & Fe & 36,866 Opteron & Jaguar & 0.000001 & $ 1.23 \times 10^{-6}$  \\
SNAP ML-IAP \cite{10.1145/3458817.3487400}  & SNAP & C & 27,900 V100 & Summit & 20 & $ {28.9} $ \\
DeePMD-kit \cite{10.1145/3503221.3508425}  & DeepPotential & Cu &  27,360 V100 & Summit & 3.4 & $9.09 $ \\
DeePMD-kit \cite{10.1145/3503221.3508425}  & DeepPotential & Cu & 9,936 A64FX & Fugaku & 1.1 & $1.49 $ \\
DeePMD-kit \cite{10.1145/3503221.3508425}  & DeepPotential & H\textsubscript{2}O &  27,360 V100 & Summit & 3.9 & $11.2 $  \\
DeePMD-kit \cite{10.1145/3503221.3508425} & DeepPotential & H\textsubscript{2}O & 9,936 A64FX & Fugaku (part) & 1.5 & $2.17 $ \\
Allegro \cite{10.1145/3581784.3627041}  & Allegro & HIV & 5,120 A100 & Perlmutter & 0.1 & $12.7 $ \\
DeePMD-kit \cite{10880101}  & DeepPotential & Cu & 90,000 SW26010 & New Sunway & 13 & $6.67$ \\
DeePMD-kit \cite{10880101}  & DeepPotential & H\textsubscript{2}O & 90,000 SW26010 & New Sunway & {29} & ${11.4}$ \\
TensorMD \cite{10.1145/3710848.3710882}  & TensorMD & W & 16,000 AMD GPU & ORISE & 4 & ${11.6}$ \\
TensorMD \cite{10.1145/3710848.3710882}  & TensorMD & W & 32,000 SW26010 & New Sunway & {51.8} & ${13.0}$  \\
GPUMD \cite{MLPAlloys2024} & UNEP-v1 & Cu & 8 A100 & cluster & 0.1 & ${0.15}$  \\
SMC-X \cite{doi:10.1021/acs.jctc.5c01614} & qSRO & HEA & 32 H800 & cluster & 128 & $1.17$  \\
\midrule
SMC-AI (This work)  & MLPNet & HEA & 32 A100 & cluster &{32} & ${0.17}$ \\
SMC-AI (This work)  & qSRO & HEA & 32 A100 & cluster &{32} & ${0.35}$ \\
SMC-AI (This work)  & qSRO & HEA & 2,048 NPU & cluster & \textcolor{red}{4096} & $\textcolor{red}{37.7 }$  \\
\bottomrule
\end{tabular} 
\end{table*} 

\section{Background}

\subsection{SMC-X}
Monte Carlo (MC) and molecular dynamics (MD) are the twin pillars of atomistic simulation, each with unique advantages in simulating equilibrium and non-equilibrium system, respectively. While MD naturally fits for large-scale parallelism, MC simulations have historically faced challenges due to the sequential nature Markov chain \cite{PREIS20094468}. Although simple models such as Ising \cite{PREIS20094468, 10.1145/3295500.3356149, ROMERO2020107473} and Lennard-Jones \cite{mick2013gpu} can exploit parallelism via checkerboard algorithms and their extensions, general short-range interactions, especially those involving ML models, lack scalable parallel MC methods. The SPMC method \cite{sadigh2012scalable}, implemented in the widely used LAMMPS package \cite{THOMPSON2022108171}, represents one of the few solutions that allows parallel MC simulations for arbitrary short-range interactions. However, SPMC’s parallelism is fundamentally constrained by its reliance on static domain decomposition, which requires large spatial domains and limits the scalability beyond one million atoms \cite{2020NPJ_Li, Yin2021}, as explicitly illustrated in the summary table of relevant works in \cite{LiuNPJ2025}.

\begin{figure}[h]
  \centering
  \includegraphics[width=\linewidth]{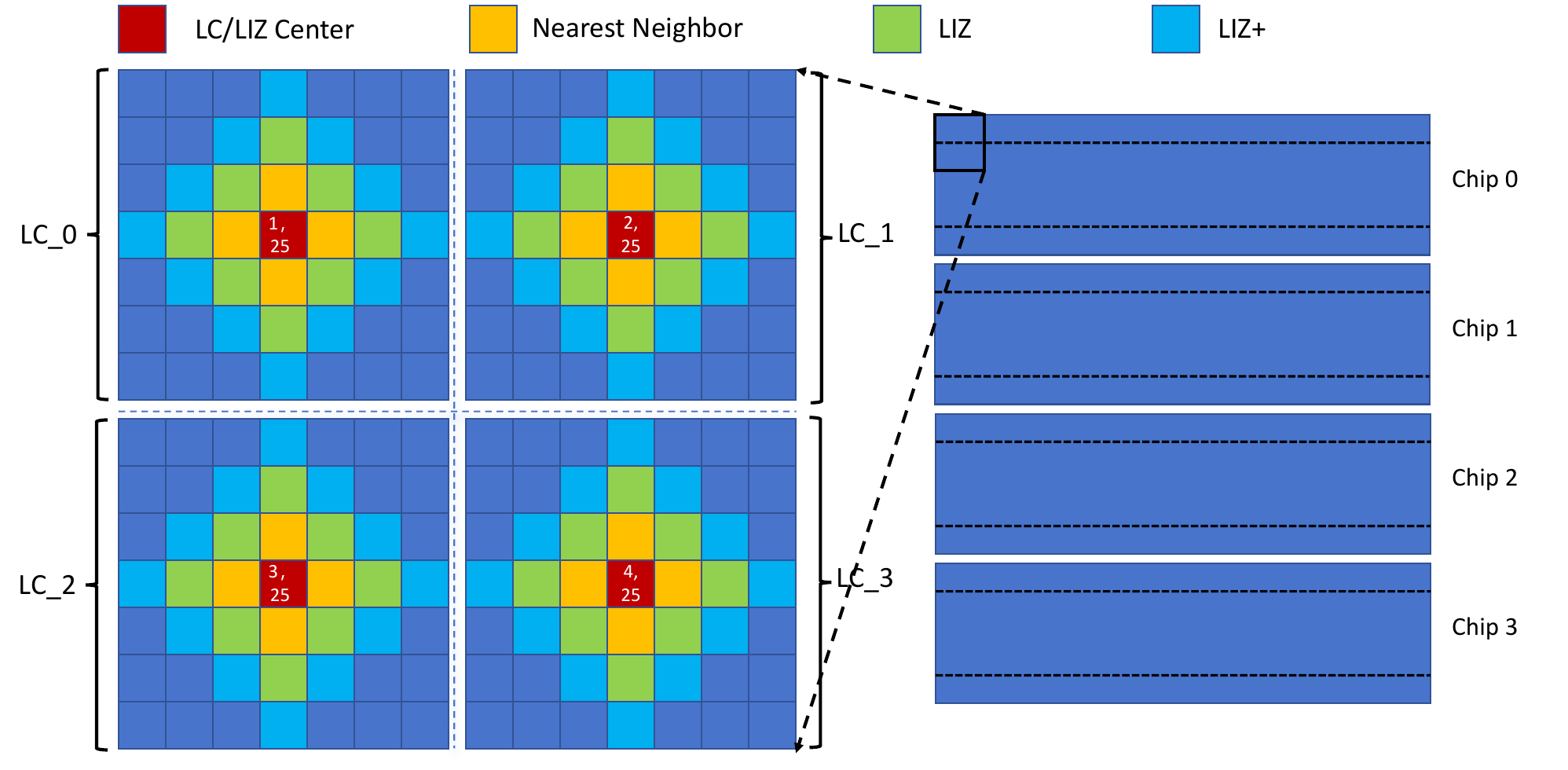}
  \caption{Schematics to illustrate the SMC-X algorithm. Adapted from \cite{LiuNPJ2025}} \label{fig:SMC_algorithm}
\end{figure}

The recently introduced SMC-X (Scalable Monte Carlo at eXtreme) method \cite{LiuNPJ2025} represents a breakthrough for massively scalable Monte Carlo simulations of arbitrary short-range interactions, enabled by several innovations illustrated in Fig.~\ref{fig:SMC_algorithm}. First, it employs a dynamic link-cell (LC) decomposition scheme: at each mini-step ($N_{LC}$ mini-step forms a complete MC step), the entire lattice is repartitioned into optimally sized link cells such that the updated atom always resides at its cell's center. This not only satisfies the detailed-balance condition crucial for simulation fidelity, but also achieves substantially higher parallelism than SPMC \cite{sadigh2012scalable}. Second, SMC-X introduces local interaction zones (LIZ) to efficiently handle non-pairwise interactions. Energy changes from Monte Carlo trials are computed through localized post-trial energy evaluations, leveraging fine-grained parallelism and data locality for significant speedups. Performance is further enhanced by implementation techniques such as on-the-fly calculation to dramatically reduce memory usage, and tight coupling of simulation and ML model evaluation on accelerator's memory hierarchy to avoid unnecessary data movement. With these innovations, SMC-X has demonstrated per-chip performance \cite{LiuNPJ2025} that exceeds all prior works \cite{doi:10.1021/acs.jctc.5c01614}, as shown in Tab.~\ref{Tab:Performance}.

\subsection{Challenges} \label{challenges}
\begin{figure}[htbp]
  \centering
  \includegraphics[width= \linewidth]{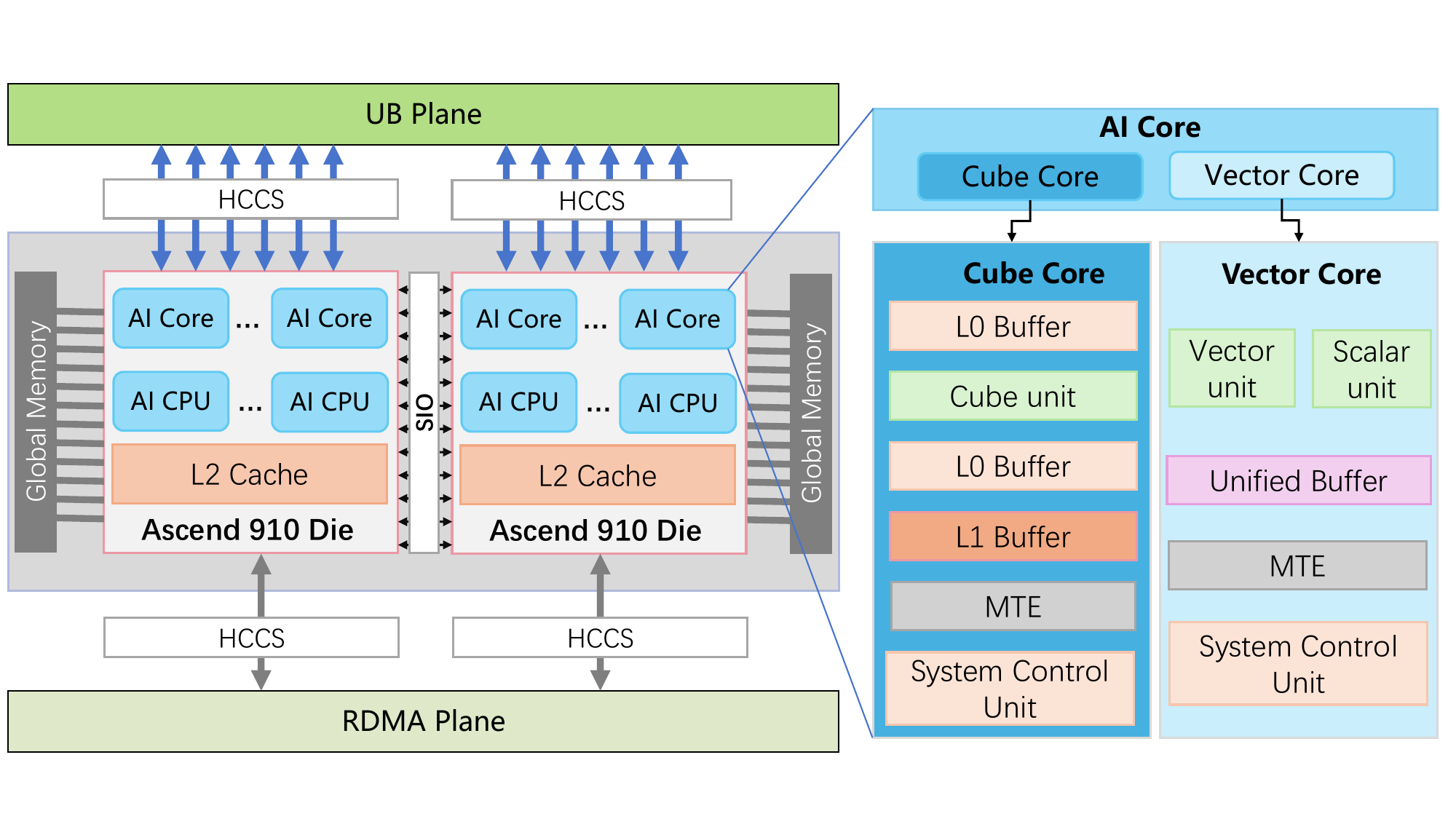}
  \caption{ A schematic of the NPU architecture. There are many AI cores in a single Ascend NPU die. Each AI core contains both Cube cores and vector cores. Adapted from  ~\cite{lin2024fastattentionextendflashattention2npus, zuo2025servinglargelanguagemodels}.} \label{fig:NPU_arch}
\end{figure}

As mentioned in the introduction, directly porting SMC-X to domain-specific AI accelerators, such as NPU,  faces severe difficulties. As shown in Fig.~\ref{fig:NPU_arch}, NPU is designed entirely for deep learning workloads, with several defining architecture characteristics: It has long vector units and large matrix units, namely the 2048-bit AI vector cores (AIV) and $16\times16\times16$ 3D AI cube cores (AIC); The memory hierarchy is designed for contiguous large-chunk data access, via simple buffers instead of sophisticated cache; The programming instruction sets eliminate branching to maximize throughput. All these features, while optimized for the efficiency of neural network training and inference, create obstacles for general HPC applications, including SMC-X, which requires fine-grained parallelism, frequent branching, and primarily irregular memory access patterns \cite{doi:10.1021/acs.jctc.5c01614}.
For reference, we found that a straightforward implementation of the {\texttt{cal\_energy()} }kernel, which is the central piece of SMC-X, takes 83 seconds per run, whereas the same kernel takes only 0.01 seconds on an A800 GPU. As will be shown, addressing this challenge requires both algorithmic innovation and optimizations tailored for NPU.

\subsection{Other related work}
To the best of our knowledge, there is currently no method that leverages domain-specific AI accelerators to enable Monte Carlo simulations with general machine-learning energy models. Existing efforts have primarily focused on accelerating Monte Carlo simulations of Ising models using classical checkerboard algorithms on AI hardware, such as TPUs and GPU tensor cores \cite{10.1145/3295500.3356149, ROMERO2020107473}. However, these approaches rely on simple nearest-neighbor pairwise interactions, limiting their applicability to idealized systems rather than realistic materials. In the context of molecular dynamics, Ref.~\cite{10793165} demonstrates significant acceleration on the Cerebras Wafer-Scale Engine compared to GPUs. Nevertheless, the model employed in that work is based on the classical Embedded Atom Method (EAM), rather than a machine-learning model.

\section{Method}
\subsection{SMC-AI} \label{SMC-AI}
\begin{figure*}[htbp]
  \centering
  \includegraphics[width=0.8 \linewidth]{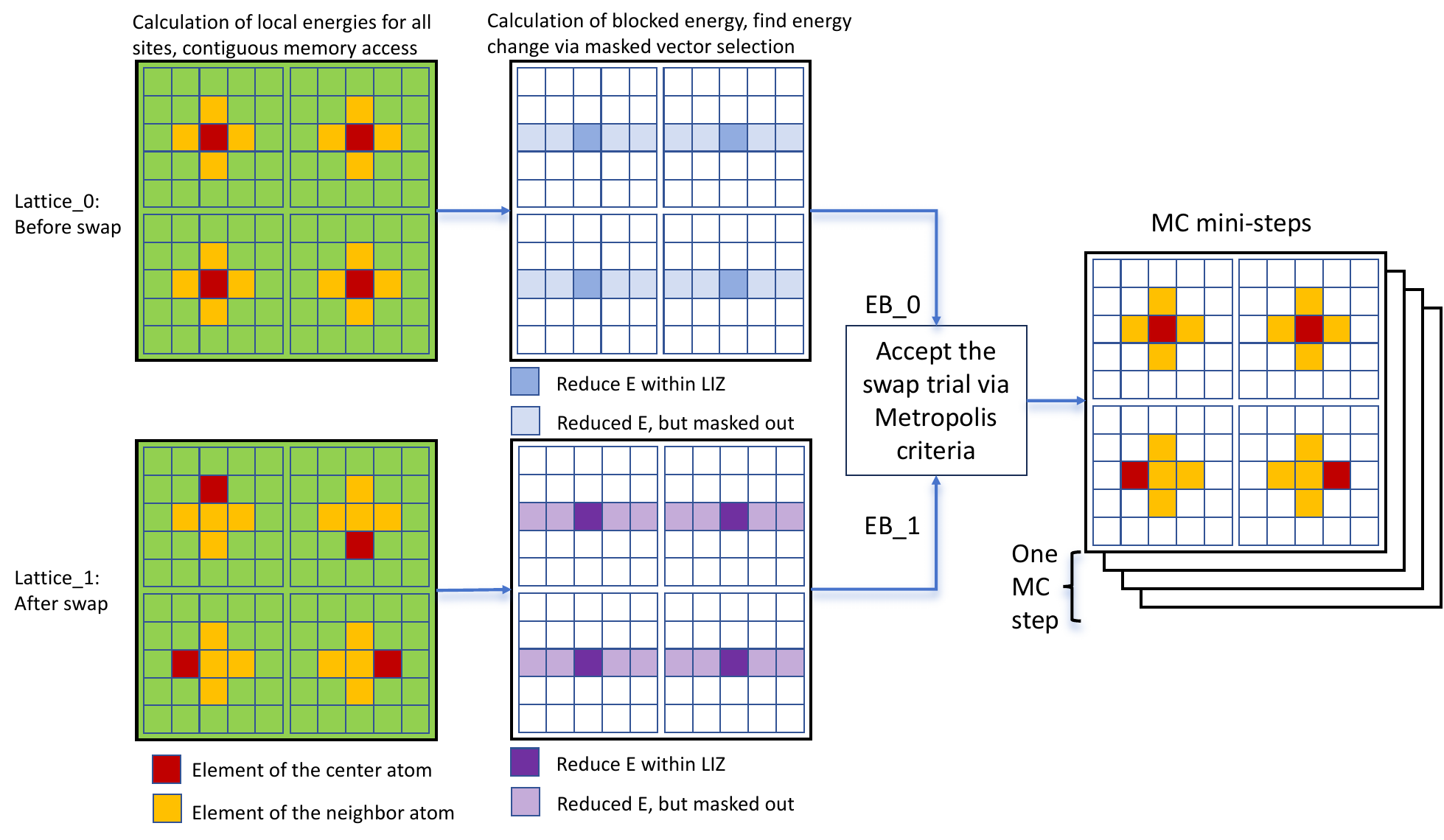}
  \caption{Schematic of a MC mini-step in a SMC-AI algorithm. To achieve \textbf{contiguous memory access} necessary for NPU, two lattices are used to store the information before and after the swap moves, with the final configurations determined by Metropolis criteria, with the help of calculated masks to make selections in vector operations.} \label{fig:SMC-AI-schematic}
\end{figure*}

To address these challenges, we introduce SMC-AI, a novel algorithm designed to fully exploit the computational power of AI chips for MC simulations while preserving the advantages of the SMC-X algorithm. The pseudocode for SMC-AI is presented in Algorithm~\ref{algorithm}, and a schematic of this workflow is provided in Fig.~\ref{fig:SMC-AI-schematic}. Note that all the steps within the for loop in Algorithm~\ref{algorithm} are embarrassingly parallel assuming no lattice decomposition. The obstacles for the efficient utilization domain-specific AI chips are eliminated by introducing a double-lattice strategy, which employs an auxiliary lattice to store post-trial atomistic configurations at each mini-step, as shown in lines 3 and 5 of Algorithm~\ref{algorithm}. For energy evaluation, the ML model is required only in the \texttt{cal\_local\_energy()} function, which is called twice per MC mini-step, as indicated in lines 4 and 6 of Algorithm~\ref{algorithm}. Moreover, by evaluating all sites using \texttt{cal\_local\_energy()}, the aforementioned contiguous memory access can be achieved, at the cost of some redundant computation, introducing a factor of approximately 2.
The energy changes resulting from the MC trial moves are computed by reducing over the affected neighboring sites, as shown in lines 7 and 8. The updated lattice is then generated from the two lattices, $\boldsymbol{\sigma}_0$ and $\boldsymbol{\sigma}_1$, using the Metropolis criterion, and overwrites $\boldsymbol{\sigma}_0$. 

Here, we present a theoretical analysis of the strengths and limitations of SMC-AI in comparison to SMC-X. First, SMC-AI eliminates irregular memory access patterns by replacing them with large, contiguous memory accesses. This transformation is critical for fully exploiting NPU architectures and avoiding the orders-of-magnitude slowdown discussed in Sec.~\ref{challenges}. 
Second, by decoupling ML model evaluation from the Monte Carlo (MC) simulation, SMC-AI significantly improves flexibility and maintainability. This design allows modifications to the ML model (e.g., switching to a different model) without impacting the MC core, thereby simplifying updates and enabling future extensions.
Furthermore, within \texttt{cal\_local\_energy()}, lattice data access and ML model evaluation are further decoupled, enabling SMC-AI to interface directly with external ML models, such as those implemented in PyTorch, once lattice information is provided by the MC core. Although such integration is beyond the scope of this work, this capability represents a significant advantage, allowing the reuse of powerful pre-trained models and facilitating the development of more accurate and efficient simulation frameworks in the future.

Finally, most of the advantages of SMC-X are retained, including algorithmic correctness from the strict satisfaction of the detailed-balance condition, high parallelism due to the elimination of sequential dependence by exploiting locality of the ML model, and exceptional scalability. The only trade-off compared to SMC-X is an approximately doubled computational effort due to redundant computation, and doubled memory footprint due to the usage of two lattices. We think this is generally an acceptable price, 
especially considering that SMC-AI remains substantially faster than nearly all other atomistic simulation methods, as shown in Tab.~\ref{Tab:Performance}, as well as the software design advantage of decoupling MC simulation with ML model inference.

\begin{algorithm}
\renewcommand{\algorithmicrequire}{\textbf{Input:}} 
 \caption{The key part of the SMC-AI algorithm. All steps within the for loop are parallelized.}\label{algorithm}
	\begin{algorithmic}[1]
	\REQUIRE The number of LCs $N_C$; the LC size $N_{LC}$; lattice $\boldsymbol{\sigma}_0(i_C, i_A)$.   
		\FOR{$i_{A}=1, 2, ..., n_A$ (\textit{sequential mini-steps})} 
            \STATE Partition $\boldsymbol{\sigma}_0$ into LCs $S({i_A})$ centered at site $i_{A}$ \\
            \STATE $\boldsymbol{\sigma}_1 = \boldsymbol{\sigma}_0$ 
            \STATE Calculate local energies $\boldsymbol{E}_0$ 
            \STATE Randomly swap trial with NN $i_N$ in $\boldsymbol{\sigma}_1$  
            \STATE Calculate local energies $\boldsymbol{E}_1$  
		      \STATE Reduce $\textbf{E}_1$ within $\Omega(i_C)$ to obtain $\textbf{E}^B_1(i_C)$
                \STATE Reduce $\textbf{E}_0$ within $\Omega(i_C)$ to obtain $\textbf{E}^B_0(i_C)$
		      \STATE Compare $\textbf{E}^B_0(i_C)$ and $\textbf{E}^B_1(i_C)$ via Metropolis criteria to decide whether accept
                \IF{\text{Accept}}
                    \STATE $\boldsymbol{\sigma}_0(i_C, i_N)$ = $\boldsymbol{\sigma}_1(i_C, i_N) $ 
                \ENDIF
		\ENDFOR
	\end{algorithmic} 
\end{algorithm}

\subsection{Implementation techniques for NPU}
Even with the same algorithm, fully exploiting the performance of AI accelerators requires tailored implementation techniques that account for the specific architecture and hardware constraints of the chips. For example, while the cube cores on NPU support the INT8 datatype, the vector cores on NPUs do not. As a result, atomic species must be stored as INT16 instead, which inevitably further increases memory usage. As another example, there is currently no high-quality random number generators (RNGs) in the NPU-CANN software stack, therefore the large amount of random numbers needed in MC have to be calculated using Huawei Kungpeng 920 CPU in the platform, in stark constrast to the GPU implementation, where the cuRAND package is used for RNG. In the following, we present other key techniques we employ to overcome the architectural limitations of the NPU. Note that many techniques can be applied to other domain-specific AI chips.

\subsubsection{Vectorization via mask}
Despite efforts in SMC-AI to minimize conditional branches, they are still inevitable for some key operations, such as element selection. For NPU, these operations are implemented via masked vectors: The vector processing unit provides a \texttt{select} operation, which enables element-wise selection between two vectors of the source operand. The result is written to a destination vector, and the operation is controlled by a mask, allowing conditional execution on a per-element basis:
\begin{equation}
    C = \text{Select(A, B, mask)}
\end{equation}
The mask perform the following:
\begin{itemize}
    \item When a mask bit is 1, the corresponding element is taken from the first source operand vector.
    \item When a mask bit is 0, the element is selected from the second source operand vector.
\end{itemize}
Note that some masks are generated by multiple steps with other auxiliary quantities. For instance, the neighbor-swap mask is obtained via bit-wise AND operations between a mask for swap-neighbor selection, and a mask for identifying $i_A$ index handled in each sequential mini-step.

\subsubsection{Hardware–software parallelism mapping}
The parallelism in SMC-X spans multiple algorithmic dimensions, including the energy model, LIZ, LC, and lattice. In contrast, the NPU computing platform offers hardware parallelism through vector/cube units, AI cores, dies/chips, and compute nodes. 
While GPUs typically provide a more implicitly managed programming model through the abstraction of 3D threads and thread blocks in CUDA, NPUs require explicit management of vector and cube units as well as AI cores. As a result, the mapping strategy adopted in SMC-AI differs significantly from that in SMC-X \cite{LiuNPJ2025}. In SMC-AI, the LIZ and LC degrees of parallelism are treated on the same footing since both correspond to local energy calculations. The total degree of parallelism available in these calculations scales linearly with the number of atoms in the lattice. At the NPU die level, this high degree of parallelism is exploited through both the large number of AIVs and the wide 2048-lane SIMD units. In the mapping strategy, the data are evenly partitioned into multiple chunks and distributed across AIVs and SIMD lanes for parallel execution.

\subsubsection{Memory Latency Hiding}
\begin{figure}[htbp]
  \centering
  \includegraphics[width=\linewidth]{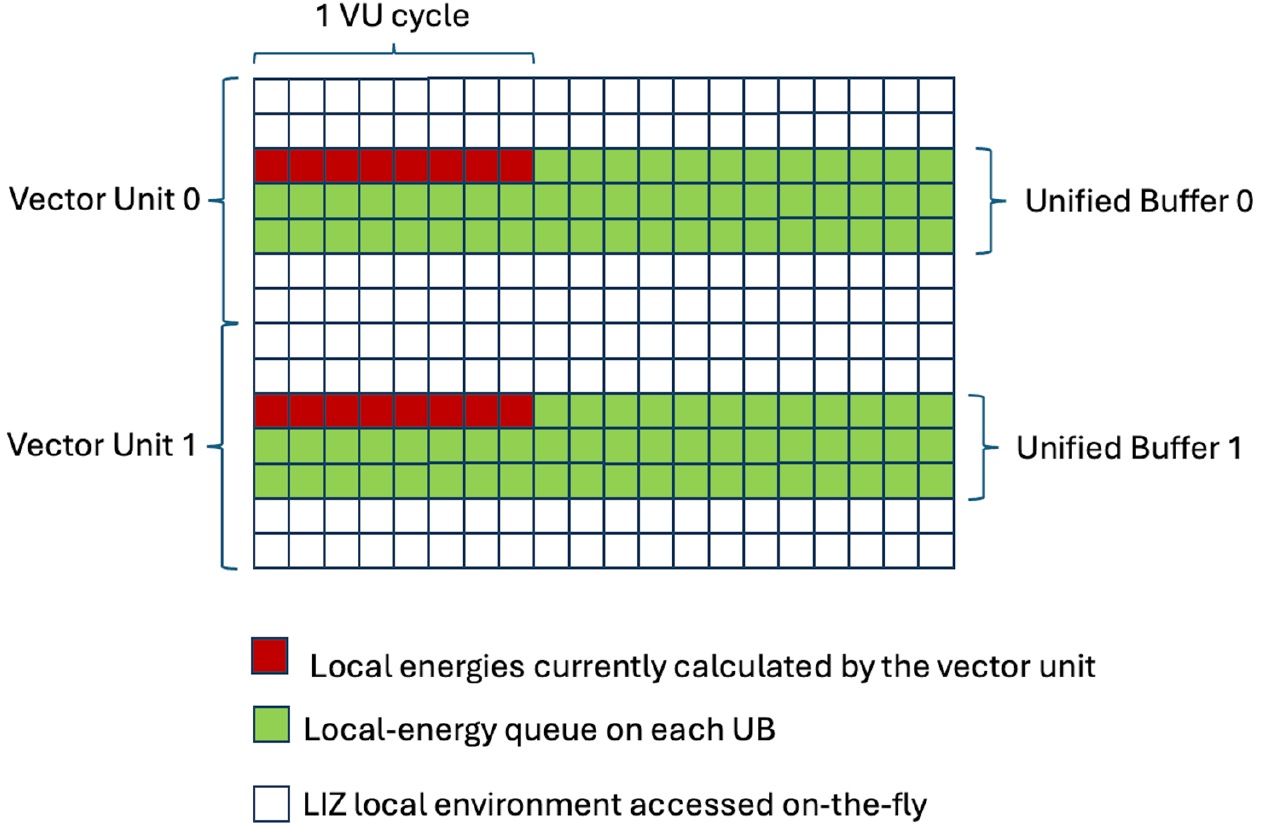}
  \caption{A schematic to illustrate the calculation of local energies via vectorization in SMC-AI. a queue of contiguous data for SIMD operation are stored in the unified buffer to hide the memory latency.} \label{fig:cal_energy}
\end{figure}

A critical distinction between GPU and NPU architectures lies in their cache hierarchy. As illustrated in Fig.~\ref{fig:NPU_arch}, NPUs lack an automatically managed L1 cache comparable to that of GPU streaming multiprocessors (SMs), necessitating explicit data movement and optimization to hide HBM latency. This is particularly important for the local-energy kernel, which constitutes the primary computational hotspot. In practice, latency hiding is achieved by prefetching queued data into the unified buffer (UB) of the AI cores, as illustrated in Fig.~\ref{fig:cal_energy} for the 2D square lattice case. For a 2048-bit AIV, 64 FP32 local energies can be computed simultaneously. To maximize buffer unit (BU) utilization and effectively hide memory latency, we carefully determine the workload assigned to each AIV. It is important to note that the input features representing the local chemical environment of each atom are computed on the fly. Therefore, compared to other atomistic simulation methods, SMC-AI inherits the key advantage of SMC-X: a substantially reduced HBM memory footprint, along with high computational efficiency achieved by minimizing unnecessary data movement and improving data locality, as demonstrated in the per chip performance of system size and simulation throughput, shown in Tab.~\ref{Tab:Performance}.

\subsubsection{PBC ghost layer}
Another key difference between SMC-AI and SMC-X lies in how periodic boundary conditions (PBC) are handled in the lattice. On NPUs, an atomic virtual layer is introduced to manage atomic exchanges across periodic boundaries, which is a standard technique for simulating crystalline systems. Note that GPUs do not require such a layer, as they can efficiently determine whether an exchanged atom lies across the block boundary and directly access the corresponding data.
However, on Ascend processors, the SIMD (Single Instruction, Multiple Data) architecture makes such conditional checks highly inefficient. As a result, the atomic virtual layer becomes necessary, even when the entire lattice supercell resides on a single accelerator. This virtual layer must be updated after each mini-step to maintain data consistency and ensure correct computation across periodic boundaries.

\subsubsection{Lattice decomposition} \label{Decomposition}
The complete lattice is distributed over all the accelerators, and implemented via 1D decomposition over the outermost $x$ dimension of the lattice. 
In weak scaling, if each MPI rank maintains a fixed local lattice shape, the communication time scales as $O(1)$. 
However, if the global lattice remains cubic, as is typical in real applications, then the communication time generally scales as $O(N^{2/3})$, and can increase dramatically once the local domain size approaches the halo thickness. 
In theory, both scenarios are less favorable than a 3D decomposition, which maintains $O(1)$ communication time under weak scaling. 
Nevertheless, 1D decomposition offers two key advantages. First, it reduces communication latency by minimizing the number of message exchanges, which is particularly beneficial for the latency-dominated regime with relatively small message sizes.
Second, it simplifies implementation and accommodates arbitrary processor counts without complex partitioning. 
For these reasons, we adopt 1D decomposition in this work. 
As demonstrated in a previous work \cite{doi:10.1021/acs.jctc.5c01614}, a simple implementation of computation--communication overlap can hide most communication overhead.

\section{Experimental Setup}
\subsection{Hardware and system software} \label{hardware}
The NPU experiments are performed on a Huawei Ascend 910 NPU cluster with similar setup as in Ref.~\cite{zuo2025servinglargelanguagemodels}. Each computational node is equiped with 2 Kunpeng 920 CPUs and 4 Ascend 910 NPUs. As illustrated in Fig.~\ref{fig:NPU_arch}, an NPU contains 2 dies, and each die contains dozens of AI cores, each one with 
1 AI Cube (AIC) core, optimized for matrix and convolution workloads, and 2 AI Vector (AIV) cores for element-wise operations. The Ascend 910 package integrates a total of 128 GB of on-package memory (64 GB per die). The package delivers up to 3.2 TB/s of aggregate memory bandwidth, with 1.6 TB/s available per die \cite{zuo2025servinglargelanguagemodels}. For interconnect, each Ascend 910 die interfaces with two-tier network fabrics. The first one (UB Plane) is for scale-up and the second one (RDMA Plane) for scale-out, with performance \cite{xiao2026huaweicloudmodelasaservicecloudmatrix384, zuo2025servinglargelanguagemodels} similar to the following H800 platform. The version of the Ascend Hardware Development Kit (HDK) used is 24.1.rc3.5, and the version of the Compute Architecture for Neural Networks (CANN) suite is 8.1.RC1. The compiler used for NPU programming is \text{BiShengCompiler-202.1.0}.

The GPU experiments are performed on commercial clusters equipped with NVIDIA H800 or A100 GPUs. The H800 platform consists of 4 nodes, each node is equipped with 2 Intel Xeon Platinum 8462Y CPUs and 8 NVIDIA H800 SXM5 GPUs. Each GPU is equipped with 80 GB of HBM3 memory and provides a theoretical memory bandwidth of 3.35 TB/s. Intra-node GPU connectivity is facilitated via NVLink, offering 400 GBps of bidirectional bandwidth, while inter-node communication is achieved through a 4× 400 Gbps InfiniBand network.
The A800 platform uses NVIDIA A800 SXM4 GPUs, each with 80 GB of HBM2 memory and a theoretical bandwidth of 2.0 TB/s, connected via a 2× 400 Gb/s InfiniBand network. The GPU computations were performed using CUDA Toolkit 12.8.

\subsection{Model}
We adopt two ML models, qSRO and MLPNet for the prediction of the configurational energies of different atomistic configurations. The qSRO model was introduced in  ~\cite{LiuNPJ2025} as a representation of general machine learning models. In qSRO, the local energy can be written as:
\begin{align}
    {\color{black}E_i = \sum_{p'<p,m} V_m^{pp'} \pi_m^{pp'} + \sum_{p,m} W_m^p (\pi_m^{pp})^2 + \lambda \sum_{p,m} (W_m^p)^2+\mathrm{C},}
\label{quadratic_local_SRO}
\end{align}
in which 
\begin{align}
    \pi_m^{pp'} = \frac{n_m^{pp'}}{\sum_{pp'} n_m^{pp'}}
\end{align}
is the percentage of $pp'$ interactions for the $m$'s coordination shell, and $n_m^{pp'}$ is the number of pairs with index $(m,p,p')$. Note that we limit the quadratic interaction to the same-element pairs to reduce the number of higher-order terms. The third term on the right of Eq.~\ref{quadratic_local_SRO} is a $l^2$ regularization of the weights of the quadratic term, encoding our prior that higher-order interactions should constitute only minor corrections.








In addition to qSRO, we develop a custom neural network model, referred to as MLPNet. The architecture consists of an input layer, followed by two hidden layers with a fixed dimension of 96, and a final regression layer. Nonlinearity is introduced using Rectified Linear Unit (ReLU) activations applied after the input layer and each hidden layer. The proposed model can be defined recursively as follows:

\begin{align}
    \mathbf{h}_1 &= \sigma(\mathbf{W}_0 \mathbf{x} + \mathbf{b}_0) \\
    \mathbf{h}_{i+1} &= \sigma(\mathbf{W}_i \mathbf{h}_i + \mathbf{b}_i), \quad i \in \{1, 2\} \\
    f(\mathbf{x}) &= \mathbf{W}_3 \mathbf{h}_3 + b_3,
\end{align}
where $\mathbf{W}_i$ and $\mathbf{b}_i$ denote the weight matrices and bias vectors of the $i$-th layer respectively, and $\sigma(\cdot)$ represents the ReLU activation function.

\begin{figure} [htbp]
    \centering
    \includegraphics[width=0.9\linewidth]{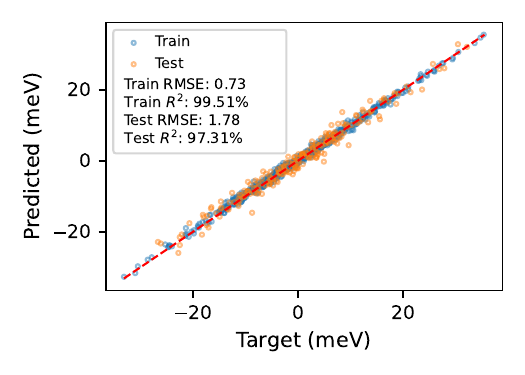}
    \caption{Parity plot of the predicted vs. DFT-calculated configurational energies for $\rm{Fe_{29}Co_{29}Ni_{28}Al_{7}Ti_{7}}$ using the MLPNet. The solid diagonal red line represents perfect agreement ($\rm{y=x}$).}
    \label{fig:parity}
\end{figure}

The ML models are trained on DFT data generated using the locally self-consistent multiple scattering (LSMS) method \cite{osti_1420087, PhysRevLett.75.2867}. Model parameters are optimized using stochastic gradient descent (SGD) with backpropagation. The EPI model achieves a test error of 2.4 meV \cite{LiuNPJ2025}, while the qSRO model yields 2.2 meV \cite{LiuNPJ2025}. In comparison, MLPNet attains a lower test error of 1.78 meV, as shown in Fig.~\ref{fig:parity}. This improved accuracy indicates that its increased model complexity enables a more accurate description of energy differences across various atomistic configurations.

\subsection{Physical system}
While the method has broad applicability \cite{zhou2026computationalmicroscopechemicalorderdisorder}, in this work we focus on the $\rm{Fe_{29}Co_{29}Ni_{28}Al_7Ti_7}$ system. This high-entropy alloy (HEA) \cite{LIU2023101018} has attracted significant attention due to its exceptional strength and ductility \cite{Yang933}, which have been attributed to the formation of coherent nanoparticles \cite{Yang933, EGeorge_Nature, 2025NatureNano}. Both the nanoparticles (solute) and the matrix (solution) adopt the fcc lattice structure, but differ in chemical ordering: the matrix is fully disordered due to entropic effects, whereas the nanoparticles form an ordered $L1_2$ structure driven by the enthalpic contribution to the free energy. The lattice constant is 6.72 Bohr (0.356 nm).
The microstructures in $\rm{Fe_{29}Co_{29}Ni_{28}Al_7Ti_7}$ and its subsystems span tens of nanometers up to 1 $\mu$m \cite{Yang933, ORNL_HEAs_2021}. Capturing these features requires system sizes at or beyond this scale, corresponding to billions of atoms. For example, a cubic system of 1 $\mu$m on a side contains roughly 88 billion atoms.

\section{Results}

\subsection{Performance}
\begin{figure}[htbp]
  \centering
  \includegraphics[width=\linewidth]{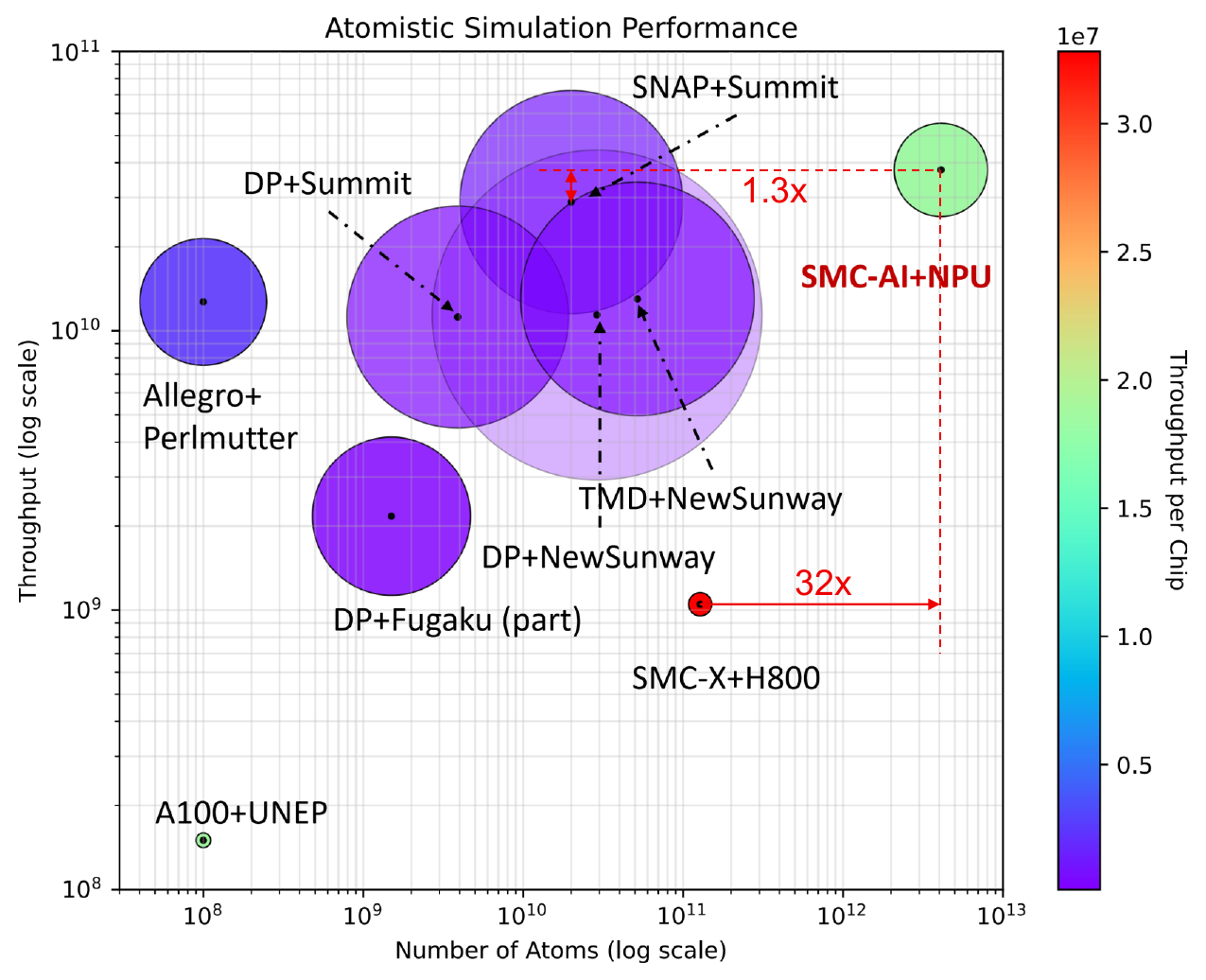}
  \caption{Comparison of the performance of our results with previous SOTA results, in terms of total $N_{atom}$ ($x$-axis),  total throughput ($y$-axis), throughput per chip (color), and the number of chips, with circles with radius proportional to $N_{Chip}^{1/3}$. Adapted from  ~\cite{doi:10.1021/acs.jctc.5c01614}.} \label{fig:performance}
\end{figure}

We compare the performance of our SMC-AI implementations with several state-of-the-art (SOTA) methods that can reach DFT-level accuracy, including MC+DFT, MD+ML, and MC+ML, as shown in Fig.~\ref{fig:performance}. Note that the throughput are measured over 10 steps, and we generally observe a one-percent statistical variation in performance, which does not affect our discussion.
This comparison should be interpreted as a reference, given the intrinsic differences between MC and MD. Nevertheless, we are striving to ensure reasonable comparability between the two approaches. For instance, we define a ``step'' in MC as one sweep, in which exactly one trial move is attempted per site, analogous to a timestep in MD. From a computational perspective, a sweep in MC is actually more expensive than an MD timestep, since many local energy evaluations are needed to determine the energy change due to a single MC swap move. Nevertheless, such comparisons between MC and MD are common in the discussions of high-performance atomistic simulation \cite{HMC_NPJ_2018}.

First, note that all ML-accelerated approaches exhibit a dramatic speed advantage compared to the DFT-based WL-LSMS method (2009 Gordon Bell Prize winner, first row in Table~\ref{Tab:Performance}). The throughput per chip is at least {${10^6}$} times higher, and typically around \textbf{$10^7$} times faster, well demonstrate the advantage of ML-accelerated atomistic simulation. It is important to emphasize that WL-LSMS is itself a highly efficient, linear-scaling DFT method, enabled by approximating electronic interactions within a finite spatial region. As a result, its scaling behavior is already comparable to that of short-range ML surrogates, and far superior to conventional cubic-scaling DFT. Thus, the orders-of-magnitude advantage of ML-accelerated approaches arises not from improved scaling, but from bypassing explicit treatment of electronic degrees of freedom.

Second, the SMC-AI implementations retain most of the performance advantages of SMC-X, consistent with the theoretical analysis in Sec.~\ref{SMC-AI}, and consistently outperform most other atomistic simulation applications in per-chip performance. Specifically, in terms of per chip performance, the NPU implementation of SMC-AI achieves \(1.84 \times 10^7\) atom\(\cdot\)step/s, while the A100 GPU implementation reaches \(1.1 \times 10^7\) atom\(\cdot\)step/s for the qSRO model. The only applications exceeding this performance are SMC-X on H800 GPUs and GPUMD on A100 GPUs; the former is expected due to the additional computational cost in SMC-AI, as discussed in Sec.~\ref{SMC-AI}, while the latter reflects the high level of GPU-specific optimization in GPUMD. When power consumption is taken into account, the performance gap between SMC-X on GPUs and SMC-AI on NPUs narrows, although GPUs still maintain a higher throughput per watt. The measured power consumption for SMC-AI on a single NPU chip (two dies) is 276~W, compared to 380~W for SMC-X on an H800 GPU, corresponding to \(6.66 \times 10^4\) atom\(\cdot\)step/J for the NPU and \(8.65 \times 10^4\) atom\(\cdot\)step/J for the GPU.

Finally, as shown in Table~\ref{Tab:Performance} and Fig.~\ref{fig:performance}, SMC-AI on 2,048 NPUs achieves the largest general-purpose atomistic simulation to date at DFT accuracy, in terms of both system size and simulation throughput. Specifically, SMC-AI reaches four trillion atoms, approximately 32$\times$ larger than the previous record of 128 billion atoms achieved by SMC-X, and about 80$\times$ larger than the largest MD simulation reported on the New Sunway supercomputer. In terms of total throughput, SMC-AI on NPUs achieves \(3.77\times10^{10}\) atom\(\cdot\)step/s, which is about 30\% of the current world record achieved on the Summit supercomputer, while using only 7.3\% of the number of accelerators (2,048 vs.\ 27,900 V100 GPUs).

\subsection{Scaling Behaviors} \label{scaling}
\begin{figure}[htbp]
  \centering
  \includegraphics[width=\linewidth]{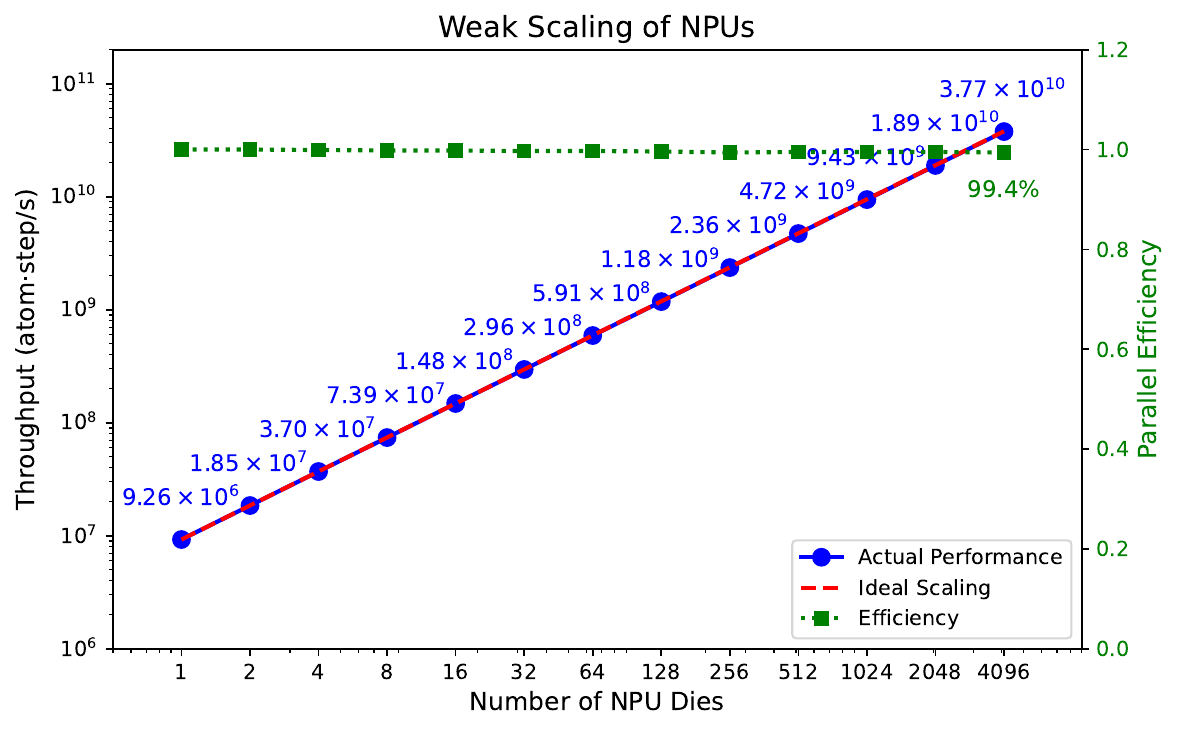}
  \includegraphics[width=\linewidth]{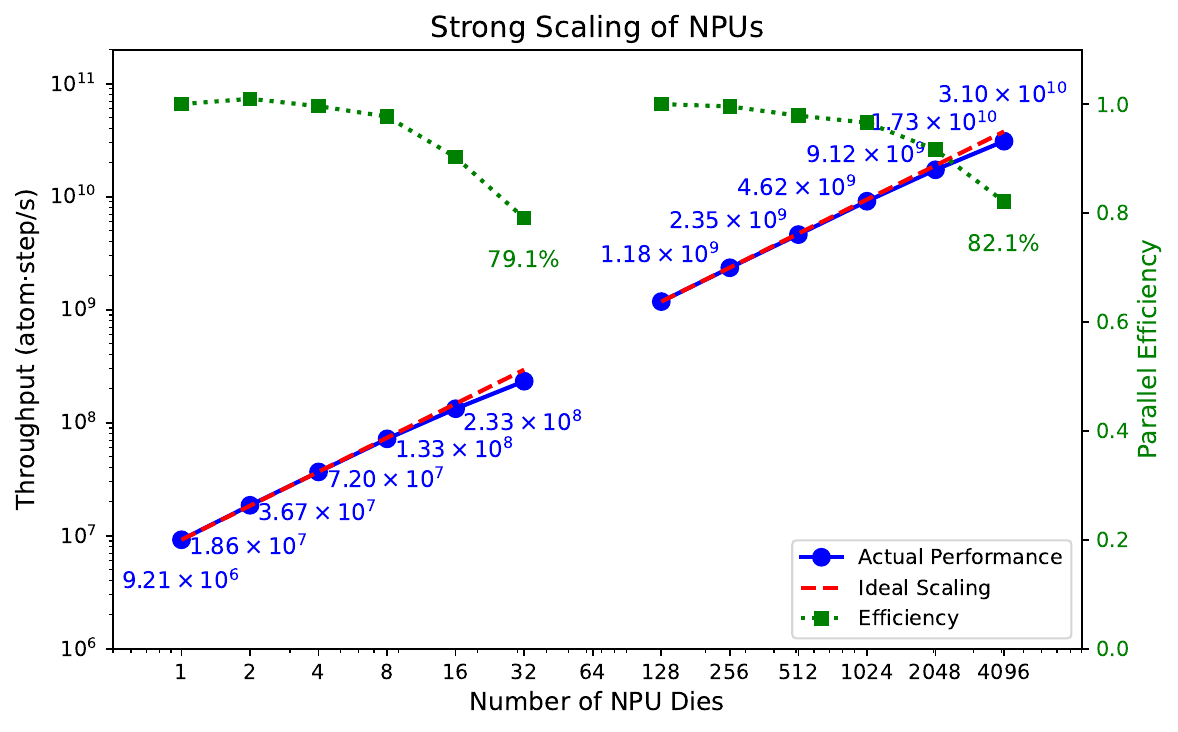}
  \caption{ The scaling behaviors of SMC-AI for NPUs. } \label{fig:scaling_NPU}
\end{figure}

\begin{figure}[htbp]
  \centering
  \includegraphics[width=\linewidth]{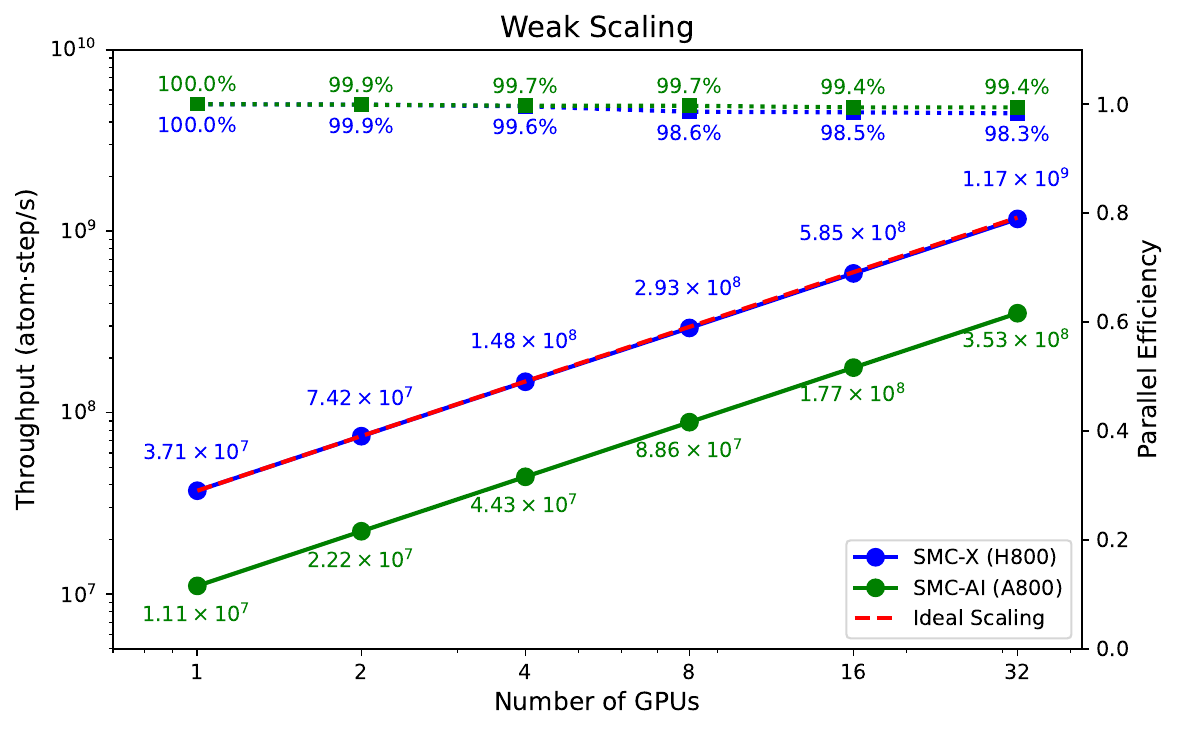}
  \includegraphics[width=\linewidth]{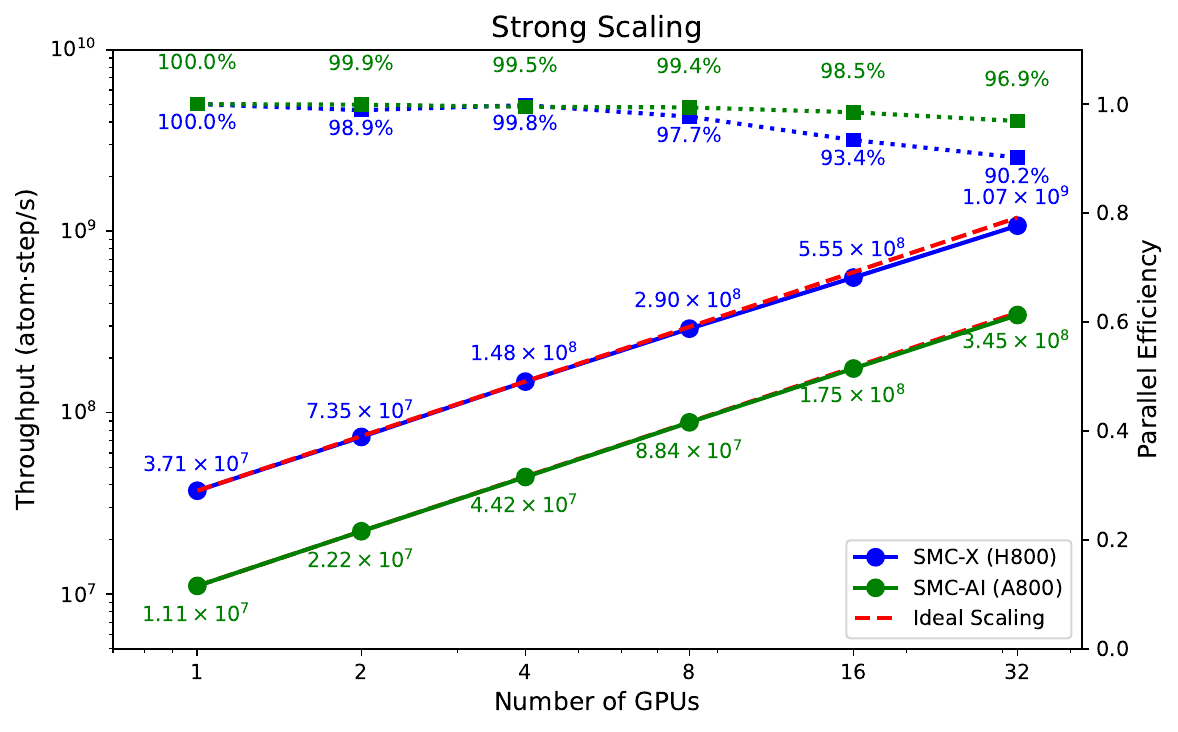}
  \caption{ The scaling behaviors of SMC-AI for GPUs. For comparison, the results for SMC-X are also listed. Note the different GPUs used.} \label{fig:scaling_GPU}
\end{figure}

The strong and weak scaling performance of SMC-AI on NPUs and GPUs are shown in Fig.~\ref{fig:scaling_NPU} and Fig.~\ref{fig:scaling_GPU}, respectively. For NPUs, two strong-scaling regimes are evaluated: a system of $1152\times468\times468\times4 = 1.009$ billion atoms is scaled up to 32 NPU dies, and a larger system of 128 billion atoms is scaled from 128 to 4096 NPU dies, achieving strong-scaling efficiencies of 79\% and 82\%, respectively. For GPUs, the smaller 1 billion-atom system is evaluated up to 32 GPUs, along with results from SMC-X, where SMC-AI achieves a strong-scaling efficiency of 90.2\%. Overall, SMC-AI demonstrates consistently strong scalability across both NPUs and GPUs, maintaining high parallel efficiency over a wide range of system sizes and hardware scales. On the other hand, both NPUs and GPUs exhibit near-ideal weak scaling, reaching 99.4\% efficiency. SMC-X generally shows lower scaling efficiency, which can be explained by its lower computational intensity compared to SMC-AI.
These results highlight the robustness of the algorithm in delivering efficient performance from moderate to extreme-scale configurations.


We further analyze the strong-scaling degradation observed on NPUs as compared to GPU.
First, the presence of ghost layers introduces additional overhead: as the domain decomposition becomes finer, the relative cost of ghost-layer computation increases, reducing overall efficiency. Second, the workload is distributed across dozens of AI vector (AIV) units, each of which requires a sufficiently large per-kernel workload to achieve high utilization. As the problem is partitioned more aggressively, the per-AIV workload decreases, leading to reduced computational efficiency and degraded scalability. Finally, the communication overhead on NPU is double of that of GPU, due to the usage of INT16 rather than INT8 for storing the chemical species at each site. 

\subsection{Communication overhead}
While the above results demonstrate that both the strong and weak scaling of SMC-AI are good, it is useful to present an analysis of the communication overhead. Consider a system with dimensions $n_x \times n_y \times n_z \times m$, where $n_x$, $n_y$, and $n_z$ represent repetitions of unit cells and $m$ denotes atoms per unit cell. The size of the communication message $M$ is given by:
\begin{equation}
M = k  \times n_y \times n_z \times n_{\text{byte}},
\label{eq:MessageSize}
\end{equation}
where $n_{\text{byte}} = 1$ (bytes) for GPU, and  $n_{\text{byte}} = 2$ for NPU, due to the different data types for storing the lattice configuration information, as aforementioned. $k$ represents the number of ghost layers required. The parameter $k$ varies with the $i_A$ index, where surface-proximal atoms require more communication layers to maintain lattice configuration consistency. For the $4\times4\times4$ link cell, we have the averaged value $k=3$. Note that the message need to be communicated for $m$ times since the four distinct fcc unit cell atoms are stored separately in our implementation.

\begin{figure}[htbp]
  \centering
  \includegraphics[width=\linewidth]{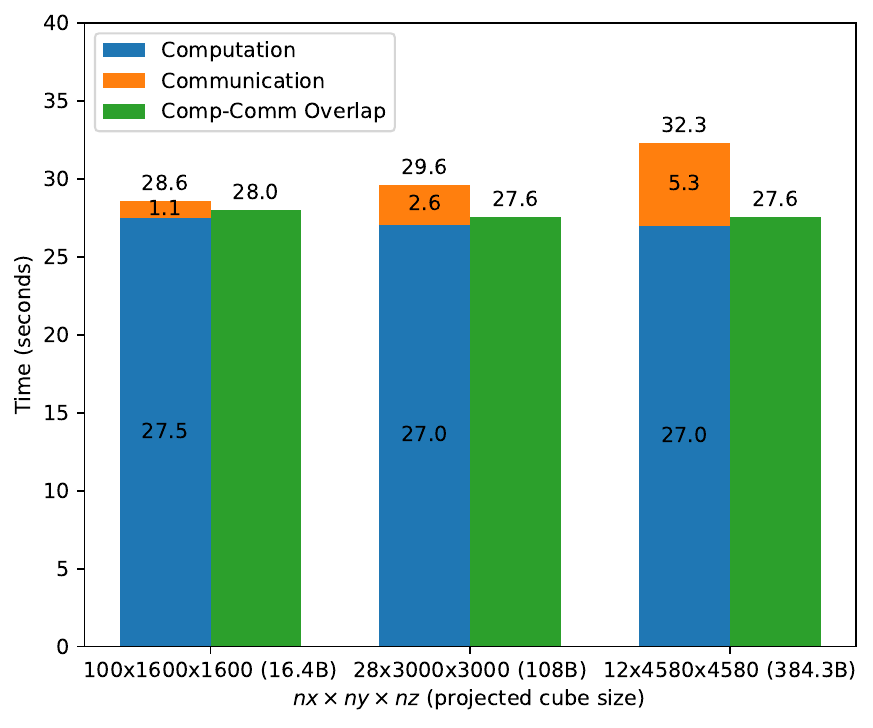}
  \caption{Effect of computation and communication overlap for different lattice shape. the $x$ label is $n_x\times n_y \times n_z$ per chip, and the parenthesis contains the total number of atoms would be for a cubic system. For each system, left bar represents total run time for non-overlap implementation, and the right one represents the overlap one.} \label{fig:overlap}
\end{figure}

While the results in section~\ref{scaling} demonstrate the excellent weak-scaling behavior of both SMC-X and SMC-AI, it should be noted that the $n_x$ dimension will increase with the number of chips. However, real systems typically prefer the cubic shape for simulation, in such a case the communication message size will increase with system size, as discussed in section~\ref{Decomposition}, 
and would become the bottom-neck, particularly for large systems of more than hundreds of billions of atoms. To address this issue, we implement computation–communication overlap and a comparison with the non-overlap results is shown in Fig.~\ref{fig:overlap}. It can be seen that for a cubic system of 16.4B atoms, communication only accounts for 3.8 percent of the total running time. However, as the system increases to 108B and 384.3B atoms, the share of time on communication increases to 8.8 percent and 16.4 percent, respectively. Note that for 16.4B, 108B, and 384.3B systems, the message sizes are 7.3 MB, 25.7 MB, and 60 MB, respectively, according to Eq.~\ref{eq:MessageSize}. In one MC step, there are $n_C=4 \times4\times 4\times4=256$ mini-steps for the $4\times4\times4$ LC and fcc lattice ( $m=4$). In each mini-step there are two ghost layers to be communicated for each rank, which give 2 send and 2 receive operations. Using the communication time information in Fig.~\ref{fig:overlap}, then the effective bidirectional bandwidth reached in each case is 27.1 GB/s, 39.9 GB/s, and 45.2 GB/s, respectively. Note that these values are much lower than the theoretical limit of the network for both IB and NVLINK (400 GB/s for the H800 system). The reason is that we did not explicitly barrier the processes before communication, for performance consideration. As an verification, applying barrier synchronization, we also measured the communication bandwidth between the 400 GB/s NVLINK of two H800 GPUs, the results are 256.7 GB/s, 291.9 GB/s, and 300.9 GB/s. As shown in Fig.~\ref{fig:overlap}, the {communication latency can be almost completely hidden} by our implementation of computation-communication overlap, even assuming cubic lattice shape in weak scaling. 

\subsection{Kernel analysis}
Here we present an analysis of the arithmetic intensity of the hotspot kernel \texttt{cal\_local\_energy}, which consists of two main steps: (1) computing the array indices of neighboring sites and reading the corresponding chemical elements, and (2) evaluating the local energies using the neighbor information. In the first step, neighbor indices and associated feature indices are computed on-the-fly via a series of \text{INT32} arithmetic operations, including addition, multiplication, and modulus, while the chemical species of neighboring sites are read from HBM, stored as \text{INT8} on GPUs and \text{INT16} on NPUs. In the second step, the local energy evaluation leverages the vector and cube units of the hardware, with performance dependent on the specific ML model employed.

It should be noted that, unlike step two which depends on the ML model, step one is an intrinsic feature of the SMC-X family of methods, including both SMC-X and SMC-AI, and therefore warrants a more detailed analysis. The neighbor index calculation primarily involves integer operations. Analysis of the kernel implementations yields a theoretical arithmetic intensity of roughly \(30\ \mathrm{IOPs/Byte}\) for GPUs and 24.8~\(\mathrm{OPs/Byte}\) for NPUs. By comparison, the hardware arithmetic intensity of the H800 GPU is approximately \(15/3.35 \approx 4.5\ \mathrm{IOPs/Byte}\), based on 15T INT32 IOPS and 3.35~TB/s HBM bandwidth. For the NPU, the hardware arithmetic intensity is approximately 6.25~\(\mathrm{OPs/Byte}\). This simple analysis indicates that the index computation step is compute-bound on integer arithmetic for both NPUs and GPUs used in this work.  
For verification, we measured the operations per second using the official NPU profiling tool \texttt{msprof}, which reports 7.03~TOPS in 16-bit operations, close to the theoretical 10~TOPS AIV hardware limit. This measurement also indicate that step one is compute-bound on integer arithmetic for NPU.

\subsection{Performance of MLPNet}
Finally, we present a study of MLPNet, which has a more complex architecture compared to the relatively simple qSRO model, due to its large number of matrix multiplication operations. To leverage the tensor cores, we use the \texttt{wmma::mma\_sync} API. Given the large number of small energy contributions in the reduction operation, we apply chunked reduction with rescaling to avoid loss of significance.  
Considering the complex interplay between GPUs, MC algorithms, and ML models, we also present an ablation study in Tab.~\ref{tab:smc_performance}, comparing single-GPU performance on H100 vs A100, SMC-X vs SMC-AI, and qSRO vs MLPNet. For the ML models, MLPNet achieves higher accuracy than qSRO (1.76~meV/atom vs 2.2~meV/atom). In terms of simulation throughput, SMC-AI is generally 2 times slower than SMC-X for qSRO, and 3 times slower for MLPNet.  
Regarding FLOPS, SMC-X achieves 108~TFLOPS for MLPNet on A100 (\(\approx 35\%\) of theoretical peak) and 169~TFLOPS on H100 (\(\approx 17\%\) of theoretical peak). SMC-AI achieves similar FLOPS, slightly higher at 118~TFLOPS (\(38\%\) efficiency). Increasing the model size (more layers or larger hidden dimensions) can further boost FLOPS, but this is impractical in practice, as it reduces simulation throughput without meaningful gains in test accuracy. The results also demonstrate that SMC-AI can effectively maintain the exceptional performance of the SMC-core after switching the model, which is an advantage over SMC-X as we mentioned in the introduction.

\begin{table}[htbp]
\centering
\caption{Performance comparison of SMC-X and SMC-AI for different ML models.}
\label{tab:smc_performance}
\begin{tabular}{c c c c c c}
\toprule
\textbf{Model} & \textbf{Accuracy} & \textbf{Algorithm} & \textbf{GPU} & \textbf{Throughput} & \textbf{TFLOPS} \\
 & (meV/atom) &  &  & (atom$\cdot$step/s) &  (FP16)\\
\midrule
qSRO & 2.2 & SMC-X & A100 & $2.39 \times 10^{7}$ & -- \\
     &     &       & H100 & $3.71 \times 10^{7}$ & -- \\
     &     & SMC-AI & A100 & $1.11 \times 10^{7}$ & -- \\
     &     &        & H100 & $1.87\times 10^7$ & -- \\
\midrule
MLPNet & 1.76 & SMC-X & A100 & $1.59 \times 10^{7}$ & 108 \\
       &      &       & H100 & $2.47\times10^7$ & 169 \\
       &      & SMC-AI & A100 & $5.35 \times 10^{6}$ & 118 \\
       &      &        & H100 & $7.54\times10^6$ & 167 \\
\bottomrule
\end{tabular}
\end{table}

\subsection{Simulation results}
\begin{figure}[htbp]
  \centering
  \includegraphics[width=\linewidth]{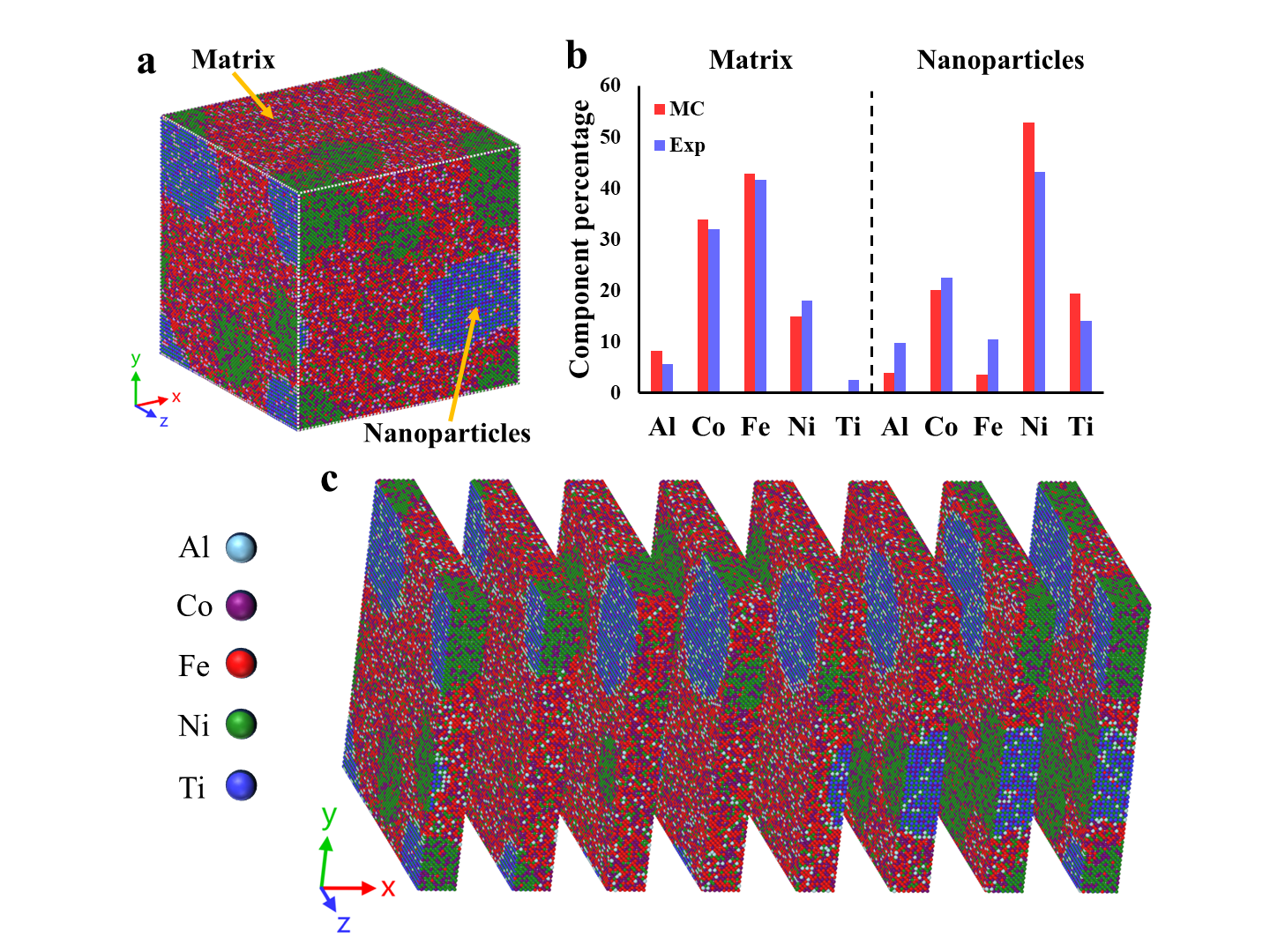}
  \caption{Simulation result for the $\rm{Fe_{29}Co_{29}Ni_{28}Al_7Ti_7}$ system. (a) 3D configurations obtained by MC simulation with qSRO model at 1000 K for 1 million MC steps, on a 1 million atom supercell. (b) Measured chemical profiles for the matrix and nanoparticle phase. (c) Sliced view of the 3D lattice.} \label{fig:1million_atom}
\end{figure}

To verify the correctness of our SMC-AI implementation, we first carry out a small-scale study using a 1-million-atom system of $\rm{Fe_{29}Co_{29}Ni_{28}Al_7Ti_7}$. The system is simulated for 1 million MC steps using the qSRO model with SMC-AI. We directly compared the results with the experimental measurements obtained by atom probe tomography \cite{Yang933}, as shown in Fig.~\ref{fig:1million_atom}.  
As observed, the system decomposes into two microphases: a disordered matrix phase composed of Fe, Co, and Ni, and $L1_2$-ordered nanoparticles, in excellent agreement with experimental observations \cite{Yang933, LiuNPJ2025}. Moreover, the chemical profiles of both the nanoparticle and matrix phases generally match the experimental data, demonstrating the correctness of the SMC-AI implementation. Our results also agree with the simulation reported in \cite{liu2025revealingnanostructureshighentropyalloys}, which employed a computationally cheaper pairwise model.

\begin{figure}[htbp]
  \centering
  \includegraphics[width=0.6\linewidth]{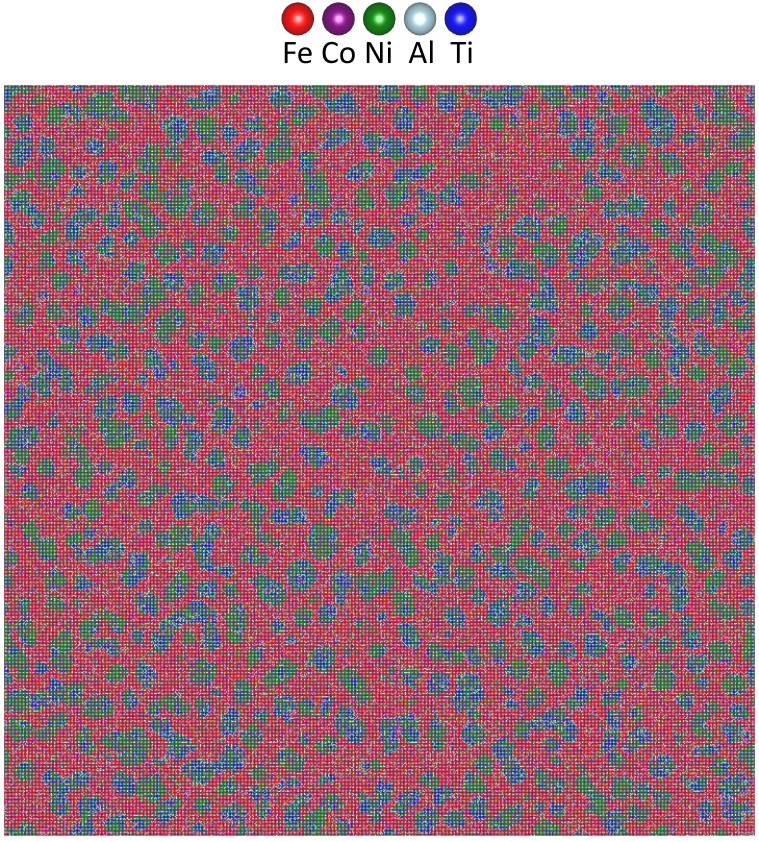}
  \caption{X-Y face of the 1 billion atom $\rm{Fe_{29}Co_{29}Ni_{28}Al_7Ti_7}$ system, obtained by MC simulation with qSRO model at 1000 K for 100,000 MC steps.} \label{fig:xyface}
\end{figure}

Finally, in Fig.~\ref{fig:xyface}, we present the MC simulation results of $\rm{Fe_{29}Co_{29}Ni_{28}Al_7Ti_7}$ on SMC-X using 16 H800 GPUs. The total system consists of 1 billion atoms. The simulation starts at 2000~K for 10,000 MC steps, then gradually decreases the temperature to 1000~K in steps of 250~K, and finally runs 100,000 MC steps at 1000~K. The total simulation time is approximately 3 days. It can be seen that the system evolves into a combination of a random solid solution phase embedded with $L1_2$ nanoparticles, in agreement with the result of 1 million atoms above, as well as the experiment \cite{Yang933}.
It should be noted the nanoparticle size observed in Fig.~\ref{fig:xyface} (~10~nm) is smaller than the experimental value of approximately 30~nm. In previous work \cite{doi:10.1021/acs.jctc.5c01614}, simulations with up to 3 million MC steps achieved better thermal equilibration, yielding nanoparticle sizes in much closer agreement with experiment, albeit using a simpler yet two-order magnitude more efficient pairwise model.

\section{Conclusion}
In this work, we introduced SMC-AI, a general algorithmic framework that extends SMC-X to enable efficient and scalable canonical Monte Carlo simulation on emerging AI accelerators. By aligning algorithmic design with the architecture of modern AI hardware, SMC-AI addresses the divergence between the HPC workload of atomistic simulation, and AI-oriented architectures. Our implementation on large-scale NPU clusters demonstrates unprecedented performance, achieving simulations of up to 4 trillion atoms on 4096 NPU dies, setting new records in both system size and throughput with strong efficiency. The portability of SMC-AI across GPU and NPU platforms further highlights its generality for next-generation heterogeneous systems. To fully exploit NPU, we employ a set of hardware-aware optimizations, including mask-based vectorization, hardware–software co-design, and memory access coalescing, enabling efficient utilization of NPUs. SMC-AI achieves strong scaling efficiencies of 82.1\% on NPUs and 96.9\% on GPUs, along with near-perfect weak scaling on both platforms. In addition, we propose the MLPNet model, which delivers higher accuracy than the qSRO model, while effectively leveraging tensor cores, with higher than 100 TFLOPS reached per GPU. The resulting simulations accurately reproduce nanoparticle formation underlying the exceptional mechanical properties of high-entropy alloys, with quantitative agreement to experimental atom probe tomography chemical profiles. These results demonstrate that SMC-AI provides an accurate, efficient, and flexible MC simulation framework that can effectively harness AI accelerators for a HPC task, therefor provide guidance on porting other HPC workloads to AI-centric hardware.

\section*{Acknowledgement}
This work was supported by Guangdong S\&T Programme under Grant 2024B0101010003. This work utilized the computing resources of Pengcheng Cloud Brain. The work of X. Liu was also supported by the National Natural Science Foundation of China under Grant 12404283. The work of F. Zhou was supported by the High
Level Talent Start-up Fund provided by Xiangnan University.
ChatGPT \cite{chatgpt_ack} has been used for polishing the language of the manuscript.

\bibliographystyle{IEEEtran}
\bibliography{sample-base}

\end{document}